\documentclass{llncs}
\usepackage[english]{babel}
\usepackage{cite}
\usepackage[utf8]{inputenc}

\usepackage{graphicx}
\usepackage{color}
\usepackage{latexsym}
\usepackage{amsfonts}
\usepackage{crayola}
\usepackage{xspace}
\usepackage{hyperref}

\title{Towards a format for describing networks}
\titlerunning{Towards a format for describing networks}
\subtitle{1. Networks and knowledge graphs}
\author{Vladimir Batagelj\inst{1,2,3,5}\orcidID{0000-0002-0240-9446} 
   \and\protect\\ Tomaž Pisanski\inst{1,2}\orcidID{0000-0002-1257-5376}
   \and Iztok Savnik\inst{1}\orcidID{0000-0002-3994-4805}
   \and\protect\\ Ana Slavec\inst{1,4}\orcidID{0000-0002-0171-2144}
   \and Nino Bašić\inst{1,2}\orcidID{0000-0002-6555-8668}
}
\authorrunning{abbreviated author list}
\institute{UP FAMNIT Koper
   \and IMFM Ljubljana
   \and UL FMF Ljubljana
   \and InnoRenew CoE Izola
   \and \email{vladimir.batagelj@fmf.uni-lj.si}
   \and \url{https://github.com/bavla/netsJSON}
}

\newcommand{\clock}{\count254=\time \divide\count254 by 60
 \count255=\count254 \multiply\count255 by -60
 \advance\count255 by \time
 \ifnum\count254<10 0\fi\number\count254\,:\,%
 \ifnum\count255<10 0\fi\number\count255}

\newcommand{\Pajek}{\texttt{\textbf{Pajek}}\xspace}

\newcommand{\keyw}[1]{\textcolor{red}{\emph{#1}}}

\newcommand{\RR}{\Bbb{R}}

\newcommand{\network}[1]{\mathcal{#1}}
\newcommand{\vertices}[1]{\mathcal{#1}}
\newcommand{\edges}[1]{\mathcal{#1}}

\newcommand{\functions}[1]{\mathcal{#1}}
\newcommand{\define}[1]{\emph{\textcolor{red}{#1}}}

\newcommand{\WA}{\mathbf{WA}}

\newcommand{\WK}{\mathbf{WK}}

\newcommand{\WC}{\mathbf{WC}}

\newcommand{\CiteN}{\mathbf{Cite}}

\newcommand{\graph}[1]{\mathcal{#1}}
\newcommand{\function}[3]{#1\,{:}\ #2\to#3}

\newcommand{\Time}{\mathcal{T}}
\newcommand{\cmdkey}{\raisebox{-.025em}{\includegraphics[height=.7em]{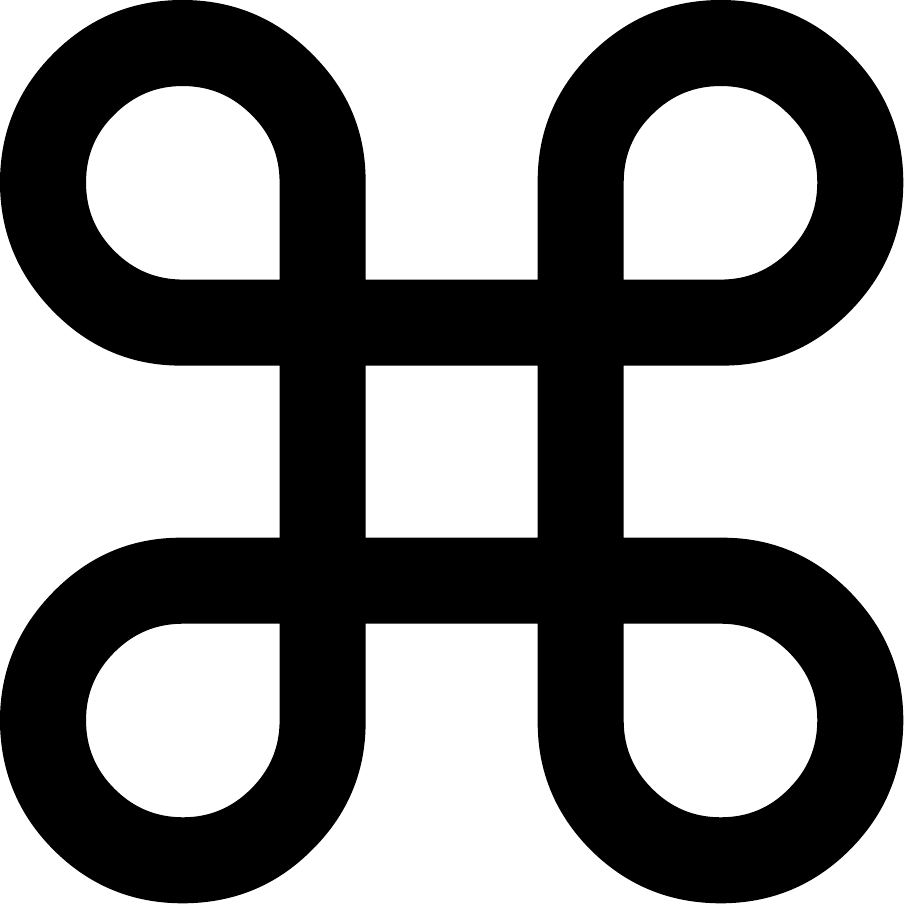}}}
\newcommand{\Mw}{\mathop{\raisebox{-1.5pt}{\mbox{$\Box$\kern-.55em\raisebox{2.5pt}{{\tiny $r$}}\kern2.9pt}}}}
\newcommand{\Mv}{\mathop{\raisebox{-1.5pt}{\mbox{$\Box$\kern-.55em\raisebox{2.5pt}{{\tiny $h$}}\kern2.9pt}}}}

\graphicspath{{./},{../pics/}}
\begin{document}
\maketitle
\begin{abstract}
The relationship between the concepts of network and knowledge graph is explored.
A knowledge graph can be considered a special type of network. When using a knowledge graph, various networks can be obtained from it, and network analysis procedures can be applied to them.
RDF is a formalization of the knowledge graph concept for the Semantic Web, but some of its solutions are also extensible to a format for describing general networks.
\end{abstract}
\keywords{Network types \and Knowledge graph \and RDF \and Identification}

\section{Introduction}

Open data plays a crucial role in ensuring the computational reproducibility and verifiability of published results. Collections of similar datasets are also crucial for developing methods to analyze specific types of data. When preparing such data, it is essential to adhere to the FAIR principles -- Findability, Accessibility, Interoperability, and Reusability. To facilitate ease of use, data should ideally be stored in a text format that preserves the structure of the data and includes relevant metadata. Datasets are alive. Their connection to open repositories is important for their accessibility and maintenance.

Network analysis is an area where data is often stored in diverse file formats. Adopting a common format for storing network data would be highly beneficial. Such a format would allow us to obtain the specific descriptions required by various network analysis programs using relatively simple scripts.


\section{Graphs and networks}

\subsection{Unit identification}

The fundamental task in transforming data about the selected topic into a structured dataset to be used in further analyses is the \keyw{identification of units} (entity recognition). Often, the source data are available as unstructured or semi-structured text. In this case, the transformation is a task of
 the \keyw{computer-assisted text analysis} (CaTA).
\keyw{Terms} considered in TA are collected in a \keyw{dictionary} (it can be fixed in advance, or built dynamically). The two main problems with terms are
\begin{itemize}
\item \keyw{equivalence} -- different words representing the same term, and
\item \keyw{ambiguity} -- same words representing different terms. 
\end{itemize}
Because of these, the \keyw{coding} -- the transformation of raw text data into formal \keyw{description} -- is often done manually or semiautomatically.

We assume that unit identification assigns a unique identifier (ID) to each unit. For some types of units, such IDs are standardized:
\href{https://en.wikipedia.org/wiki/ISO_3166-1_alpha-2}{ISO 3166-1 alpha-2} two-letter country codes,
\href{https://en.wikipedia.org/wiki/ISO_9362}{ISO 9362} Bank Identifier Codes (BIC),
\href{https://en.wikipedia.org/wiki/ORCID}{ORCID} -- Open Researcher and Contributor ID,
\href{https://en.wikipedia.org/wiki/ISSN}{ISSN} -- International Standard Serial Number,
\href{https://en.wikipedia.org/wiki/Digital_object_identifier}{DOI} -- Digital Object Identifier,
\href{https://en.wikipedia.org/wiki/Uniform_Resource_Identifier}{URI} -- Uniform Resource Identifier,
etc.

In data displays, often IDs are replaced by corresponding (short) labels/names.


Besides the semantic units or \keyw{concepts} related to the selected topic, we can identify in the raw data also syntactic units -- parts of the text. As \keyw{syntactic units} of TA we usually consider clauses, statements, paragraphs, news, messages, etc.

In thematic TA the units are coded as a rectangular matrix\break
\textit{Syntactic units} $\times$ \textit{Concepts} which can be considered as a two-mode network.


In semantic TA the units (often clauses) are encoded according
to the S-V-O (\textit{Subject}-\textit{Verb}-\textit{Object})
model or its improvements.\bigskip

 \centerline{\includegraphics[width=55mm,bb=11 7 152 47,clip=]{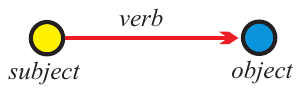}}\medskip


This coding can be directly considered as a network with
\textit{Subjects} $\cup$ \textit{Objects} as nodes and links (arcs)
labeled with \textit{Verbs}.


\subsection{Networks}

A \keyw{network} is based on two sets -- a set of \keyw{nodes} (vertices), that represent the selected \keyw{units}, and a set of \keyw{links} (lines), that represent \keyw{ties} between units. They determine a \keyw{graph}. A link can be \keyw{directed} -- an \keyw{arc}, or \keyw{undirected} -- an \keyw{edge}.

Additional data about nodes or links can be known -- their \keyw{properties} (attributes).
For example: name/label, type, value, etc.\\
\centerline{{\bf\textcolor{BrickRed}{\textbf{Network $=$ Graph $+$ Data}}}}
The data can be measured or computed.

\newpage
\subsection{Networks Formally}

A \define{network} $\network{N}=(\vertices{V},\edges{L},\functions{P},\functions{W})$
consists of:
\begin{itemize}
   \item
      a \define{graph} $\graph{G}=(\vertices{V},\edges{L})$, where $\vertices{V}$ is
      the set of nodes, $\edges{A}$ is the set of arcs,
      $\edges{E}$ is the set of edges,
      and $\edges{L}=\edges{E}\cup\edges{A}$ is the set of links. \\
       $n=|\vertices{V}|$, $m=|\vertices{L}|$
   \item
      $\functions{P}$ is a set of \define{node value functions} /
      properties: $\function{p}{\vertices{V}}{A}$
   \item
      $\functions{W}$ is a set of \define{link value functions} /
      weights: $\function{w}{\edges{L}}{B}$
\end{itemize}

Additional information/data about values:
\begin{itemize}
   \item How can we compute with values -- algebraic structures: semigroup, monoid, group, semiring, etc.
   \item Properties of values -- in a molecular graph, an atom is assigned to each node; properties of relevant atoms are such additional data.
\end{itemize}

\subsection{Some terminology}

\parbox{60mm}{\includegraphics[width=60mm]{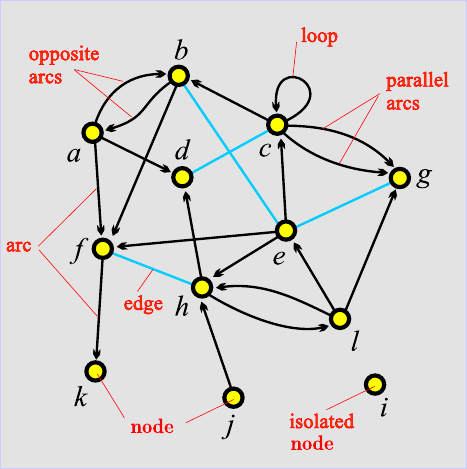}} \quad
\parbox{45mm}{unit, actor --  node, vertex \\
tie, line -- link, edge, arc\medskip\\ 
\keyw{arc} $=$ directed link, $(a,d)$\\
$a$ is the \keyw{initial} node, \\
$d$ is the \keyw{terminal} node. \medskip \\ 
\keyw{edge} $=$ undirected link, $(c\colon d)$ \\
$c$ and $d$ are \keyw{end} nodes.
}


\section{Types of networks}

Besides ordinary (directed, undirected, mixed) networks, some extended types of networks are also used:
\begin{itemize}
\item \keyw{2-mode networks}, bipartite (valued) graphs -- networks between two disjoint sets of nodes.
\item \keyw{multi-relational networks}.
\item \keyw{linked networks} and \keyw{collections of networks}.
\item \keyw{multilevel networks}.
\item \keyw{temporal networks}, dynamic graphs -- networks changing over time.
\item specialized networks:
   representation of genealogies as
   \href{http://eclectic.ss.uci.edu/~drwhite/pgraph/p-graphs.html}%
    {\textit{p-graphs}}; \keyw{Petri's nets}, etc.
\end{itemize}

Network (input) file formats should provide a means of expressing all of these types of networks.
All interesting data should be recorded (respecting privacy).

\subsection{Two-mode networks}

In a \keyw{two-mode} network $\network{N}=((\vertices{U},\vertices{V}),\edges{L},\functions{P},\functions{W})$ the set of nodes consists of two disjoint sets of nodes $\vertices{U}$ and $\vertices{V}$, and all the links from $\edges{L}$ have one end node in  $\vertices{U}$ and the other in $\vertices{V}$.
Often also a \keyw{weight} $w:\edges{L}\to\RR \in \functions{W}$ is given; if not, we assume  $w(u,v)=1$ for all $(u,v)\in \edges{L}$.

A two-mode network can also be described by a rectangular matrix  $\mathbf{A} = [a_{uv}]_{\vertices{U} \times \vertices{V}}$.
\[ a_{uv} =
 \left\{\begin{array}{ll} 
   w(u,v)  \quad & (u,v) \in \edges{L} \\
   0  & \mbox{otherwise}
 \end{array}\right. \]

Examples:
(persons, societies, years of membership),
(buyers/consumers, goods, quantity),
(parliamentarians, problems, positive vote),
(persons, journals, reading),
(authors, works, is an author).
A classical example of a two-mode network is the Southern women (Davis 1941). 
.

\subsection{Multi-relational networks}

A \keyw{multi-relational network} is denoted by
\[ \network{N} =(\vertices{V}, (\edges{L}_1,\edges{L}_2,\ldots,\edges{L}_k),\functions{P},\functions{W})\]
and contains different relations $\edges{L}_i$ (sets of links) over the same set of nodes. Also, the
weights from $\functions{W}$ are defined on different relations or their union.

Examples of such networks are the transportation system in a city (stations, lines);
\href{https://wordnet.princeton.edu/}{WordNet}
(words, semantic relations: synonymy, antonymy, hyponymy, meronymy, etc.),
\href{https://eventdata.parusanalytics.com/data.html}{KEDS} -- Kansas Event Data System networks (states, relations between states: Visit, Ask information, Warn, Expel person, etc.).


\subsection{Linked networks and collections of networks}

In a \keyw{linked} or \keyw{multimodal} network 
\[ \network{N} =((\vertices{V}_1,\vertices{V}_2,\ldots,\vertices{V}_j), (\edges{L}_1,\edges{L}_2,\ldots,\edges{L}_k),\functions{P},\functions{W})\]
the set of nodes $\vertices{V}$ is partitioned into subsets (\keyw{modes}) $\vertices{V}_i$, $\edges{L}_s \subseteq \vertices{V}_p \times \vertices{V}_q$, and properties and weights are usually partial functions.

A set of networks $\{  \network{N}_1,  \network{N}_2, \ldots,  \network{N}_k \}$ in which each network shares a set of nodes with some other network is called a \keyw{collection} of networks.

A linked network can be transformed into a collection of networks and vice versa.

Bibliographical information is usually represented as a collection of bibliographical networks $\{ \WA, \CiteN, \WK, \WC, \mathbf{WI}, \ldots \}$ (W -- works, A -- authors, K -- keywords, C -- countries, I -- institutions) \cite{batagelj2013bibliographic}.


\subsection{Multilevel networks}

A \keyw{multilevel network} organizes nodes into hierarchical levels, where each level represents a different scale or granularity of interaction or connectivity. It is a special case of a linked network.

Key characteristics of multilevel networks are:
hierarchical levels, inter-level links, scale-specific dynamics, and modularity.

Each level may have its unique dynamics, rules, or behaviors. For example, in a biological network, one level might represent protein-protein interactions, while another level represents cellular interactions.
They often exhibit modularity, where nodes within a level are more densely linked to each other than to nodes in other levels. This modularity can help in understanding the functional or structural organization of the network.

Example: a multilevel network in the context of a university: Level 1: Individual students and faculty members, Level 2: Departments or academic units, Level 3: The entire university.

\subsection{Temporal networks}

In a temporal network, the presence/activity of a node/link can change through time $\Time$.

\subsubsection{Temporal quantities.}

To describe changes, we introduce a notion of a \keyw{temporal quantity} (TQ) \cite{batagelj2016algebraic, batagelj2020temporal}
\[  a(t) = \left\{\begin{array}{ll} 
                a'(t) & t \in T_a \\
                \cmdkey & t \in \Time \setminus T_a
             \end{array}\right. \]
where $T_a$ is the \keyw{activity time set} of $a$ and $a'(t)$ is the value of $a$ in an instant $t \in T_a$, and  \cmdkey{} denotes the value \keyw{undefined}.

We assume that the values of temporal quantities belong to a set $A$ which is
a \keyw{semiring} $(A,+,\cdot,0,1)$ for binary operations $+ : A\times A \to A$ and
$\cdot : A\times A \to A$. $A_{\scriptsize\cmdkey} = A \cup \{\cmdkey\}$.

Let $A_{\scriptsize\cmdkey}(\Time)$ denote the set of all temporal quantities
over $A_{\scriptsize\cmdkey}$ in time $\Time$. To extend the operations to
networks and their matrices we first define the \keyw{sum} (parallel links)
$ a + b $  as
\[ (a+b)(t) =  a(t) + b(t) \quad \mbox{ and } \quad T_{a + b} = T_a \cup T_b .\]
The \keyw{product} (sequential links) 
$ a \cdot b $ is defined as             
\[ (a \cdot b)(t) =  a(t) \cdot b(t) \quad \mbox{ and } \quad T_{a \cdot b} = T_a \cap T_b . \]

Let us define TQs $\mathbf{0}$ and $\mathbf{1}$ with
requirements $\mathbf{0}(t) = \cmdkey$ and $\mathbf{1}(t) = 1$ for all
$t \in \Time$. Again, the structure 
$(A_{\scriptsize\cmdkey}(\Time),+,\cdot,\mathbf{0},\mathbf{1})$ is a
semiring.

To produce software support for computation with TQs, we limit it to TQs that can be described as a sequence of disjoint time intervals with a constant value 
\[ a = [(s_i,f_i,v_i)]_{i \in 1..k} \] 
where $s_i$ is the starting time and $f_i$ the finishing time of the $i$-th time interval $[s_i, f_i)$, $s_i < f_i$ and $f_i \leq s_{i+1}$, and $v_i$ is the value of $a$ on this interval.  Outside the intervals, the value of TQ $a$ is undefined, $\cmdkey$. 


For example
$a  = [(1, 5, 2), (6, 8, 1), (11, 12, 3), (14, 16, 2),  (17, 18, 5), (19, 20, 1)]$.
Assuming discrete time, the TQ $a$ has at time points 1, 2, 3, 4 value 2; it is undefined at time point 5; has at time points 6, 7 value 1; etc.

A \keyw{temporal network}
\[ \network{N}_T =(\vertices{V},\edges{L},\functions{P},\functions{W}, T)\]
is obtained if the \keyw{time} $T$ is attached to an ordinary network.
$T$ is a linearly ordered set of  \keyw{time points} $t \in T$. In a temporal network, some node properties and/or link weights are assigned TQs.

In a temporal network, nodes $v \in \vertices{V}$ and links $l \in \edges{L}$ are not necessarily present or active at all time points. If a link $l(u,v)$ is active at time point $t$ then also its end nodes $u$ and $v$ should be active in time $t$.

We will denote the network consisting of links and nodes active in time $t \in T$ by $\network{N}(t)$ and call it a \keyw{time slice} in time point $t$.

Program \Pajek\ also supports descriptions of temporal networks based on \keyw{events} -- a sequence of transformations (add, remove, change, time point, etc.).

\subsection{Multi-relational temporal network -- KEDS/WEIS}

Types of networks can be combined. Here is a fragment from the Balkans Data Set of \href{https://eventdata.parusanalytics.com/data.html}{KEDS} data
{\renewcommand{\baselinestretch}{0.7}\footnotesize
\begin{verbatim}
890402  YUG     KSV     224  (RIOT)  RIOT-TORN
890404  YUG     ETHALB  212  (ARREST PERSON) ALB ETHNIC JAILED IN YUG
890407  ALB     ETHALB  224  (RIOT)  RIOTS
890408  ETHALB  KSV     123  (INVESTIGATE)   PROBING
 ...
030731  GER     CYP     042  (ENDORSE)       GAVE SUPPORT
030731  UNWCT   BOSSER  212  (ARREST PERSON) SENTENCED TO PRISON
030731  VAT     EUR     043  (RALLY) RALLIED
030731  UNWCT   BOSSER  013  (RETRACT)       CLEARED
030731  UNWCT   BAL     121  (CRITICIZE)     CHARGES
030731  SER     UNWCT   122  (DENIGRATE)     TESTIFIED
030731  BOSSER  UNWCT   121  (CRITICIZE)     ACCUSED
...
\end{verbatim}
}

\noindent
that determines a multi-relational temporal network.


\section{Knowledge graphs}

\subsection{Knowledge graph}

A knowledge graph is a structured representation of knowledge that captures entities, relationships, and attributes in a graph-based format. New entities, relationships, and attributes can be added dynamically.
Knowledge graphs have evolved from early semantic networks (1960s) and ontologies to become powerful tools for representing and reasoning about complex knowledge. The Austrian linguist Edgar W. Schneider coined the term "knowledge graph" in 1972.

A knowledge graph is a graph of data intended to accumulate and convey knowledge of the real world, whose nodes represent entities of interest and whose edges represent potentially different relations between these entities \cite{hogan2021knowledge}. It acquires and integrates information into an ontology and applies a reasoner to derive new knowledge \cite{ehrlinger2016towards}.

Knowledge graphs are widely used in applications like semantic search, recommendation systems, natural language processing, and artificial intelligence. They are often built using standards like RDF (Resource Description Framework) and queried using languages like SPARQL \cite{RDF12draft, SPARQL12, angles2008expressive}.

\subsection{Knowledge graphs formally}

There is no generally accepted definition of a knowledge graph. Most approaches are based on the view that it is a set of \keyw{facts} $\mathcal{F}$, where each fact is a triple of the form $(e_1, r, e_2)$ or $(e, a, v)$, with:
  $e,  e_1, e_2 \in E $, 
  $ r \in R $,
  $ a \in A $, and
  $v$ (value, which can be a literal or another entity). $E$ is a set of \keyw{entities} (nodes), representing real-world objects, concepts, or instances. $ R $ is a set of \keyw{relationships} (arcs), representing directed links between entities. $ A $ is a set of \keyw{attributes}, representing properties or characteristics of entities or relationships.

\begin{figure}
\centerline{\includegraphics[width=\textwidth]{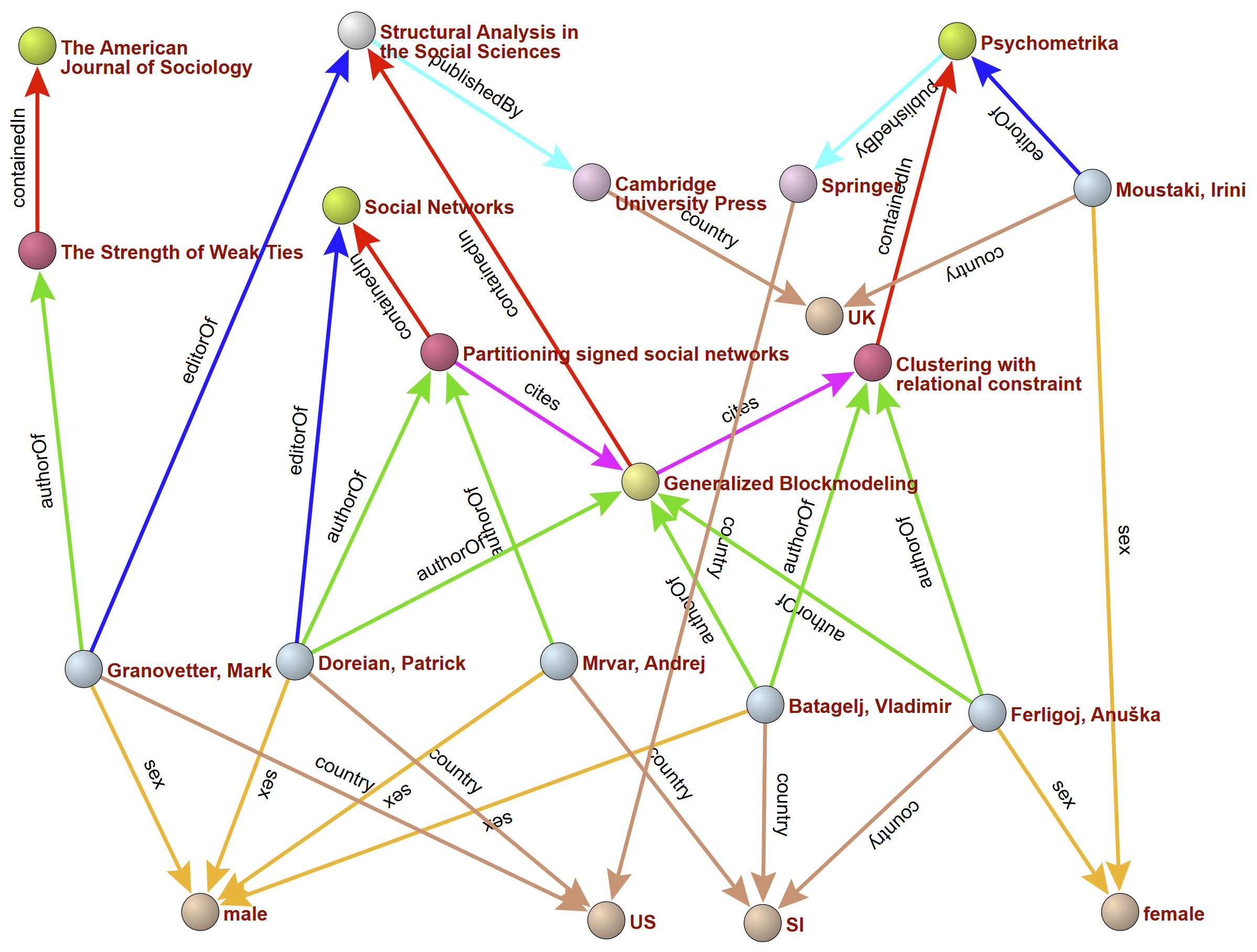}}
\caption{Bibliography as knowledge graph}
\end{figure}

The set of facts $\mathcal{F}$ can be represented with a \keyw{knowledge graph} (network)
\[ G = (\mathcal{V}, \mathcal{A}, \lambda) \]
such that for every arc $\ell(u,v) \in \mathcal{A}$ the triple $(\lambda(u), \lambda(\ell), \lambda(v)) \in \mathcal{F}$ and there is an arc in $\mathcal{A}$ for each triple from $\mathcal{F}$. There is no isolated node in $\mathcal{V}$. The function $\lambda \ \colon \mathcal{V}\cup \mathcal{A} \to \mathcal{L} $, where $ \mathcal{L} $ is a set of \keyw{labels}, provides semantic meaning to entities, relationships, and attributes.

In a network, the labels on nodes are considered standard node properties; arcs with the same label determine the corresponding relation. Therefore, the knowledge graph $G$ is a multi-relational network.
Usually, the attribute arcs and attribute value nodes are transformed into functions (node properties).

\subsubsection{Example.}


As an example, let us describe a part of the bibliography determined by the following works:
\href{http://www.cambridge.org/tw/academic/subjects/sociology/sociology-general-interest/generalized-blockmodeling}{Generalized blockmodeling}, \href{http://link.springer.com/article/10.1007%2FBF02293706}{Clustering with relational constraint}, \href{http://www.sciencedirect.com/science/article/pii/S0378873308000397}{Partitioning signed social networks}, \href{http://www.journals.uchicago.edu/doi/abs/10.1086/225469}{The Strength of Weak Ties}.

There are nodes of different types (modes): persons, papers, books, series, journals, publishers;  different relations between them:  author\_of, editor\_of, contained\_in, cites, published\_by; and attributes: sex and country. See Figure 1.

\subsection{Limitations}


An S-V-O triple on which the knowledge graph is based corresponds in logic to a binary predicate $V(S, O)$. In sentences describing facts predicates of larger -arity can appear. 

An approach would be to use hypergraphs instead of graphs. 
An alternative approach is to express a k-ary predicate as a binary predicate. Consider a 4-ary predicate $P(x,y,t,z)$. The quadruple $(x,y,t,z)$ can be expressed as a pair in three ways $(x,(y,t,z))$, $(x,y),(t,z))$, and $((x,y,t),z)$. Defining a binary predicate $P'(x,(y,t,z)) = P(x,y,t,z)$ we get the triple $(x,P',(y,t,z))$.
Now, we have to consider the triple $(y,t,z)$ as an object. This is called a reification.
In knowledge representation, the \keyw{reification}  is the process of turning a predicate or statement into an addressable object. Reification allows the representation of assertions so that they can be referred to or qualified by other assertions \cite{Reification, Nary}.

Another problem is how to express an existence fact in which an entity is not known -- we only know that it exists. In logic, this is resolved by \keyw{Skolemization}. We can deal with both problems by introducing blank nodes \cite{chen2012blank}.



\section{RDF Graphs}

RDF  is a standard model for data interchange on the web, developed by the World Wide Web Consortium (W3C). It is designed to represent information about resources in a way that can be easily processed by machines. It is a key component of the Semantic Web, which aims to make web data more meaningful and interconnected.
RDF provides a universal framework for describing and linking data, making it easier for machines to understand and process information across different systems. It plays a crucial role in enabling the vision of a more intelligent and interconnected web.

In \cite{RDF12draft} we learn ``There can be four kinds of nodes in an RDF graph: IRIs (Internationalized Resource Identifier), literals, blank nodes, and triple terms'' (see ``1.2 Resources and Statements'').
An IRI is a Unicode string for unambiguously identifying nodes and arcs. IRIs are internationalized versions of URIs, which are generalizations of URLs. A blank node is appropriate when the node does not need to be referenced directly. Blank nodes can be reached by following their incident arcs from other nodes. Literals are concrete values used to represent data.

In ``3. RDF Graphs'' a definition of an RDF graph is given:

An \keyw{RDF graph} is a set of RDF triples.
An RDF triple is said to be \keyw{asserted} in an RDF graph if it is an element of the RDF graph.
An \keyw{RDF triple} (often simply called "triple") is a 3-tuple that is defined inductively as follows:
\begin{enumerate}
\item If $s$ is an IRI or a blank node, $p$ is an IRI, and $o$ is an IRI, a blank node, or a literal, then $(s, p, o)$ is an RDF triple.
\item If $s$ is an IRI or a blank node, $p$ is an IRI, and $o$ is an RDF triple, then $(s, p, o)$ is an RDF triple.
\end{enumerate}
The three components $(s, p, o)$ of an RDF triple are respectively called the subject, predicate, and object of the triple.

Note: The definition of triple is recursive. That is, a triple can itself have an object component that is another triple. However, by this definition, cycles of triples cannot be created.

IRIs, literals, blank nodes, and triple terms are collectively known as RDF terms.

Instead of traditional RDF graph triple representation with labeled nodes and arcs \medskip

\includegraphics[width=40mm]{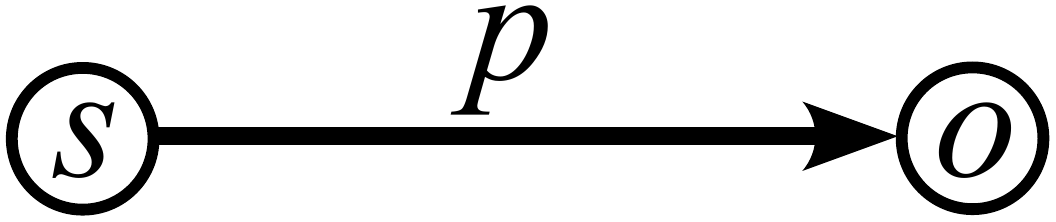} \medskip

\noindent
we propose a representation based on the following alternative \medskip

 \includegraphics[width=40mm]{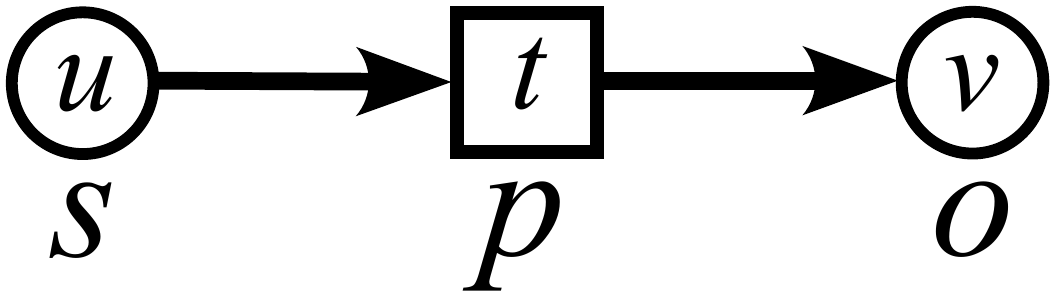} \qquad or \qquad  \includegraphics[width=40mm]{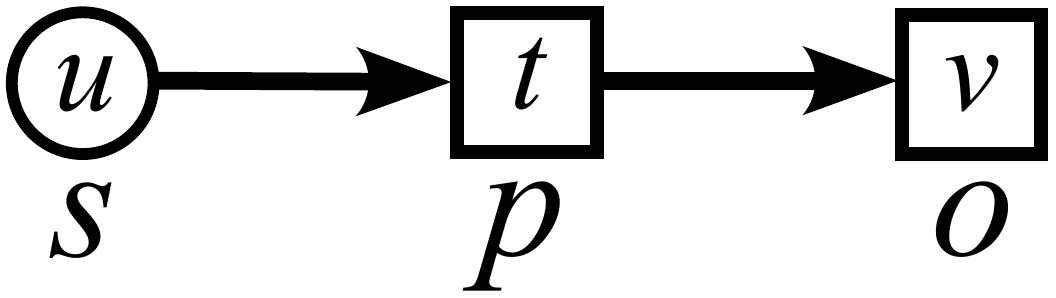}\medskip

\subsection{RDF networks}

In the proposed approach an RDF graph is a network ${\cal K} = (({\cal S},{\cal T}), {\cal A}, \lambda)$.
${\cal S}$ is the set of simple nodes, ${\cal T}$ is the set of triple nodes, and ${\cal V}= {\cal S}\cup {\cal T}$ is the set of nodes. ${\cal A}$ is the set of arcs (directed links) and $\lambda$ is the label function -- for the example triple $\lambda(u) = s$, $\lambda(v) = o$, and $\lambda(t) = p$.

Labels are from three sets ${\cal L}$ -- literals, ${\cal B} = \{ \textvisiblespace \}$ -- blank, and ${\cal I}$ -- IRI. The inductive definition of an RDF graph can be transformed as follows. Let $\bbbk$ be the set of \keyw{RDF networks}. Then

\textbf{RDFb.} Empty network ${\cal K} = ((\emptyset,\emptyset), \emptyset, \emptyset) \in \bbbk$.

Let ${\cal K} \in \bbbk$ and ${\cal K}' = (({\cal S}',{\cal T}'), {\cal A}', \lambda')$ be the expanded network. Then for every $u \in {\cal V}$ we have $\lambda'(u) = \lambda(u)$. 

\textbf{RDFs.} If $\lambda'(u) \in {\cal I}\cup {\cal B}$, $\lambda'(v) \in {\cal I}\cup {\cal B}\cup {\cal L}$, $t \in S$, and $\lambda'(t) \in {\cal I}$ then  ${\cal S}' = {\cal S}\cup \{u,v\}$, ${\cal T}' = {\cal T}\cup \{t\}$, ${\cal A}' = {\cal A}\cup \{(u,t), (t,v)\} $, and the network ${\cal K}' \in \bbbk$.

\textbf{RDFt.} If $v \in {\cal T}$, $u \in {\cal S}$, $\lambda'(u) \in {\cal I}\cup {\cal B}$ then  ${\cal S}' = {\cal S}\cup \{u \}$, ${\cal T}' = {\cal T}\cup \{t\}$, ${\cal A}' = {\cal A}\cup \{(u,t), (t,v)\} $, and the network ${\cal K}' \in \bbbk$.

Let $n_S = |{\cal S}|$, $n_T = |{\cal T}|$, and $m = |{\cal A}|$. For the initial empty network $n_S = n_T = m = 0$. For each application of the rules \textbf{RDFs} and \textbf{RDFt} we have $n_T' = n_T + 1$ and $m' = m + 2$. Therefore, after $k$ applications of these rules, it holds $n_T = k$ and $m = 2k$. 

Some interesting questions about the RDF networks:
Can a graph ${\cal K} = (({\cal S},{\cal T}), {\cal A})$ be recognized as a graph of an RDF network?
The oriented $K_4$ with $n_S = 1$ and $n_T = 3$ is not in $\bbbk$.
Can we obtain the corresponding construction sequence?

\begin{figure}
\centerline{\includegraphics[width=100mm]{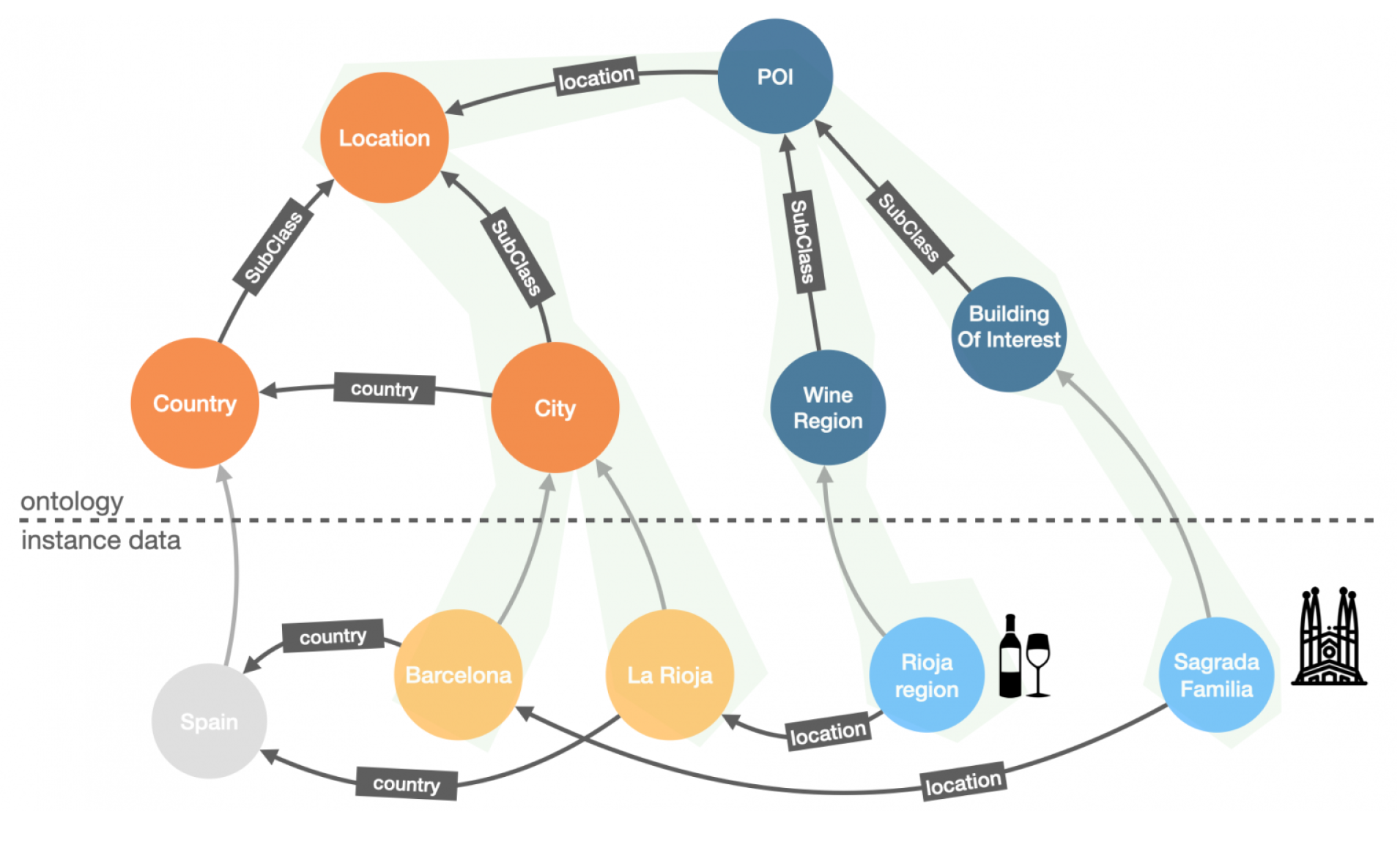}}
\caption{Barcelona \href{https://www.data-expo.nl/en/blog/what-is-a-knowledge-graph}{(Data Expo 2025)}}
\end{figure}

In applications, the set of facts of the knowledge graph is augmented by general (common) knowledge about the application field (ontology -- also expressed as triples \cite{OWL2}). See Figure 2. This allows for more effective use of logic in solving problems in a given field.

\section{Networks and analysis of knowledge graphs}

From the knowledge graph, we can create various networks, from which we can obtain useful information using network analysis procedures.

If we are only interested in the structure of the network, we can forget about the predicate/relation labels and analyze the resulting simplified network.

Another option is to obtain subnetworks of individual predicates/relations and, using network multiplication (and other operations), create meaningful derived networks, which are then analyzed.

For example, from the bibliographic knowledge graph we can obtain the networks mentioned in subsection 3.3. The product $\mathbf{Co} = \mathbf{WA}^T*\mathbf{WA}$ gives us the co-authorship network. Its entry $Co[a,b]$ counts the number of works in which $a$ and $b$ are co-authors. Similarly, $AK[a,k]$ from the derived network $\mathbf{AK} = \mathbf{WA}^T*\mathbf{WK}$ tells us in how many works author $a$ has dealt with the topic described by the keyword $k$. The derived network $\mathbf{ACi} =
\mathbf{WA}^T * \mathbf{Cite} * \mathbf{WA}$ describes citations between authors -- the entry $ACi[a,b]$ counts how many times author $a$ cited author $b$, etc. \cite{batagelj2013bibliographic}.

\section{Conclusions}

Knowledge graphs can have a very diverse structure (a large number of types of units and predicates), which allows for a fairly accurate description of facts from a selected field and solving problems about it. Network analysis methods are particularly useful in analyzing one or a few relational (sub)networks of a knowledge graph.


\subsection{Acknowledgments}
 
The computational work reported in this paper was performed using programs R  and Pajek  \cite{de2018exploratory}. The code and data are available at Github/Bavla \cite{BavlaEx}.

V.\ Batagelj is supported in part by the Slovenian Research Agency
 (research program P1-0294, research program CogniCom (0013103)
at the University of Primorska, and research projects J5-2557, J1-2481, and J5-4596)
 and prepared within the framework of the COST action CA21163 (HiTEc).

T.\ Pisanski is supported in part by the Slovenian Research Agency 
(research program P1-0294 and research projects BI-HR/23-24-012, J1-4351, and J5-4596).

N.\ Bašić is supported in part by the Slovenian Research Agency 
(research program P1-0294 and research project J5-4596).


\bibliographystyle{splncs04}

\end{document}